\documentclass[superscriptaddress,twocolumn,showpacs,pre,floatfix]{revtex4-1}

\usepackage{amsmath}
\usepackage{color}
\usepackage{tabularx}
\usepackage[dvips]{graphicx}

\begin{document}

\title{Hyperscaling violation in Ising Spin Glasses}

\author{P. H.~Lundow} \affiliation {Department of Mathematics and
  Mathematical Statistics, Ume{\aa} University, SE-901 87 Ume{\aa}, Sweden}

\author{I. A.~Campbell} \affiliation{Laboratoire Charles Coulomb
  (L2C), Univ. Montpellier, CNRS, Montpellier, France}

\begin{abstract}
  In addition to the standard scaling rules relating critical
  exponents at second order transitions, hyperscaling rules involve
  the dimension of the model.  It is well known that in canonical
  Ising models hyperscaling rules are modified above the upper
  critical dimension. It was shown by M. Schwartz in 1991 that
  hyperscaling can also break down in Ising systems with quenched
  random interactions; Random Field Ising models which are in this
  class have been intensively studied. Here numerical Ising Spin Glass
  data relating the scaling of the normalized Binder cumulant to that
  of the reduced correlation length are presented for dimensions 3, 4,
  5 and 7. Hyperscaling is clearly violated in dimensions 3 and 4 as
  well as above the upper critical dimension $D=6$. Estimates are
  obtained for the "violation of hyperscaling exponent" values in the
  various models.
\end{abstract}


\maketitle

\section{Introduction}\label{sec:I}
It has been tacitly or explicitly assumed that Edwards-Anderson Ising
Spin Glasses (ISGs), where the quenched interactions are random,
follow the same basic scaling and Universality rules as in the
canonical Ising models, whose properties are understood in great
detail through Renormalization Group Theory (RGT). Here we will
present numerical evidence for hyperscaling violation in ISGs.  A
textbook definition of hyperscaling is :"Identities obtained from the
generalised homogeneity assumption involving the space dimension D are
known as hyperscaling relations." \cite{simons:97}.  The hyperscaling
relations valid in canonical Ising models below the upper critical
dimension are : $2-\alpha = D\nu$, and $2\Delta = D\nu +\gamma$, where
$\Delta = \gamma+\beta$ is the ``gap'' exponent associated
with the critical behavior of the higher field derivatives of the free
energy \cite{butera:02,pelissetto:02}.  The two hyperscaling relations
are linked through the Essam-Fisher relation $\alpha + D\nu -
2\Delta = 2$.

Hyperscaling "collapses" in Ising models in dimensions above the upper
critical dimension $D=4$, where the critical exponents become mean
field : $\gamma=1$, $\nu =1/2$, $\alpha = 0$ and $\Delta =
3/2$. Hyperscaling was predicted by Schwartz to break down also in
quenched systems with random interactions \cite{schwartz:91}.  The
breakdown of hyperscaling in the 3D Random Field Ising model (RFIM)
has been extensively studied
\cite{aharony:76,gofman:93,vink:10,fytas:13}. The first hyperscaling
relation in this model is re-written $2-\alpha = (D-\theta)\nu$ where
$\theta$ is the ``violation of hyperscaling
exponent''~\cite{aharony:76} with $\theta \sim 1.47$ in the 3D RFIM
\cite{middleton:02,fernandez:11}.  Logically the second hyperscaling
relation should simultaneously become $2\Delta = (D-\theta)\nu
+\gamma$.

Though not conventionally written this way, in the standard Ising
models above $D=4$ equivalent modified hyperscaling relations
$2-\alpha = (D-\theta)\nu = 2$ and $2\Delta = (D-\theta)\nu +\gamma =
3$ can be seen by inspection to be consistent with the mean field
exponents plus a violation exponent $\theta =D-4$.

Ising spin glasses (ISGs) are also systems with quenched randomness in
which hyperscaling might be expected to break down, from a
generalization of Schwartz's argument. The exponent $\alpha$ in ISGs
is always strongly negative and so is very hard to estimate directly;
we will be concerned only with the second hyperscaling relation. We
are not aware of any tests of hyperscaling in ISGs.

\section{Scaling}\label{sec:II}
In numerical simulation analyses the conventional RGT based approach
consists in using as the thermal scaling variable the reduced
temperature $t = (T-T_{c})/T_{c}$, together with the principal
observables $\chi(t,L)$ the [reduced] susceptibility
$\Sigma_{x}\langle S(x).S(0)\rangle$ , $\xi(t,L)$ the second moment
correlation length, and the Binder cumulant, $g(t,L) = (3-\langle
m^4\rangle/\langle m^2\rangle^2)/2$ in Ising models, or $g(t,L) =
(3-\lbrack\langle q^4\rangle\rbrack/\lbrack\langle
q^2\rangle\rbrack^2)/2$ in spin glasses. For the simulation data the
standard finite-$L$ definition for the second-moment correlation
length $\xi(\beta,L)$ through the Fourier transformation of the
correlation function is used, see for instance
Ref.~\cite{hasenbusch:08} Eq.~(14).  The conventional approach is
tailored to the critical region; however at high temperatures $t$
diverges and $\xi(t,L)$ tends to zero, so it is not possible to
analyse the entire paramagnetic regime without introducing diverging
correction terms. In addition, in symmetric interaction ISGs the
relevant interaction strength parameter is $\langle J_{ij}^2\rangle$
so the ISG thermal scaling variable should depend on the square of the
temperature; this basic point was made some thirty years ago
\cite{singh:86,klein:91} but has been ignored since in most ISG
simulation data analyses.

A rational scaling approach which takes in the entire paramagnetic
region so including both the finite-size scaling regime (FSS, $L \ll
\xi(\beta,\infty)$) and the thermodynamic limit regime (ThL $L \gg
\xi(\beta,\infty)$), can be based on the Wegner scaling expression for
the bulk Ising susceptibility \cite {wegner:72}
\begin{equation}
  \chi(\tau)= C_{\chi}\tau^{-\gamma}[1+ a_{\chi}\tau^{\nu\omega} + b_{\chi}\tau + \cdots]
\end{equation}
where $\tau = 1-\beta/\beta_{c}$ in Ising models with $\beta$ the
inverse temperature~\cite{domb:62}.  (The Wegner expression is often
mis-quoted with $t$ replacing $\tau$). The terms inside $[\cdots]$ are
scaling corrections, with $\nu\omega$ the leading thermal correction exponent
which is identical for all observables within a universality class. At
infinite temperature $\tau=1$, and for all $S=1/2$ models $\chi(\tau)$
tends to $1$; hence for the susceptibility the whole paramagnetic
region can generally be covered to good precision when a few mild
Wegner correction terms are included. (To obtain infinite precision an
infinite number of correction terms are of course needed, just as in
standard FSS analyses where perfect precision in principle requires an
infinite number of size dependent correction terms
\cite{ferrenberg:18}). In ISG models where the interaction
distributions are symmetric about zero, an appropriate thermal scaling
variable which should be used with the same Wegner expression is $\tau
= 1-(\beta/\beta_{c})^2$,
Refs.~\cite{singh:86,klein:91,daboul:04,campbell:06}.  In the ThL
regime $L \gg \xi(\beta)$ the properties of a finite-size sample if
normalized correctly are independent of $L$ and so are the same as
those of the infinite-size model. A standard rule of thumb for the
approximate onset of the ThL regime is $L \gtrsim 7 \xi(\beta,L)$ and
the ThL regime can be easily identified in simulation data.  An
important virtue of this approach is that the ThL numerical data can
be readily dovetailed into High Temperature Series Expansion (HTSE)
values calculated from sums of exact series terms (limited in practice
to a finite number of terms). No such link can be readily made when
the conventional FSS thermal scaling variable $t$ is used.

To apply the Wegner formalism to observables $Q$ other than $\chi$ in
$S=1/2$ models, it is convenient to impose the rule that each
observable should be normalized in such a way that that the infinite
temperature limit is $Q_{n}(\tau=1) \equiv 1$ , without the critical
limit exponent being modified. For the spin $S=1/2$ [reduced]
susceptibility with the standard definition no normalization is
required as this condition is automatically fulfilled, with a
temperature dependent effective exponent $\gamma(\tau) =
\partial\ln[\chi(\tau,L)]/\partial\ln\tau$ both in Ising models, and
in ISGs with the appropriate $\tau$.

In Ref.~\cite{campbell:06} the reduced second moment correlation
length was introduced : $\xi(\tau,L)/\beta^{1/2}$ in Ising models, and
$\xi(\tau,L)/\beta$ in ISG models. The critical limit ThL exponent
$\nu$ is unaltered by this normalization (models with zero temperature
critical points are a special case \cite{lundow:16}).  From exact and
general HTSE infinite-temperature limits for either Ising or ISG
models, this reduced correlation length tends to $1$ at infinite
temperature \cite{butera:02,daboul:04}. 

The temperature dependent effective exponent is then defined as
$\nu(\tau) = \partial\ln[\xi(\tau,L)/\beta^{1/2}]/\partial\ln\tau$ in
Ising models and $\nu(\tau) =
\partial\ln[\xi(\tau,L)/\beta]/\partial\ln\tau$ in ISG models. A
Wegner-like relation again applies with appropriate correction terms;
with this definition the temperature dependent effective $\nu(\tau)$
usually turns out to remain close to the critical value, except at
high temperatures where it may be significantly modified by the
corrections.

The temperature dependent effective exponents $\gamma(\tau)$ and
$\nu(\tau)$ are well behaved over the entire paramagnetic regime with
exact infinite-temperature hypercubic-lattice limits for Ising models
of $\gamma(1) = 2D\beta_{c}$ and $\nu(1) = D\beta_{c}$, and for the
ISG models $\gamma(1) = 2D\beta_{c}^2$ and $\nu(1) =
(D-K/3)\beta_{c}^2$ where $K$ is the kurtosis of the interaction
distribution.

\begin{figure}
  \includegraphics[width=3.4in]{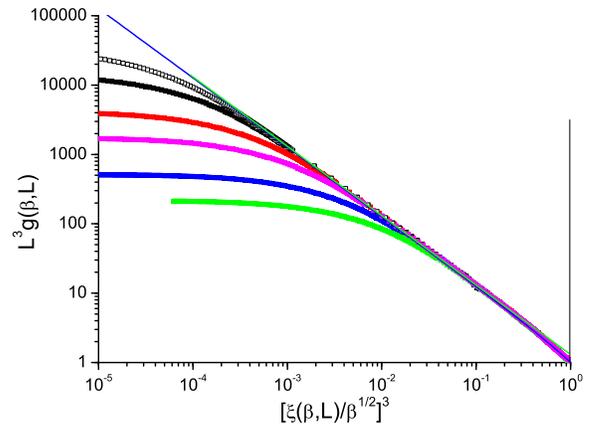}
  \caption{(Color on line) Dimension 3 simple cubic Ising
    model. Normalized Binder cumulant $L^{3}g(\beta,L)$ against
    reduced correlation length to the power 3,
    $1/(\xi(\beta,L)/\beta^{1/2})^3$. Sample sizes : $L=32$, $24$,
    $16$, $8$, $6$ (top to bottom). Blue straight line : slope
    $-1.00$. In this and all following figures each line for fixed $L$
    begins to bend over towards horizontal when it leaves the ThL
    regime $L >> \xi(\beta)$.}\protect\label{fig1}
\end{figure}

\section{Hyperscaling}\label{sec:III}
The second field derivative of the bulk susceptibility
$\chi_{4}(\beta)$ (also called the non-linear susceptibility) in a
hypercubic lattice is directly related to the ThL Binder cumulant for
finite $L$ through
\begin{equation}
  2g(\beta,L) = \frac{-\chi_{4}}{L^D\chi^{2}} = \frac{3\langle
    m^2\rangle^2 - \langle m^4\rangle}{\langle m^2\rangle^2}
\end{equation}
see Eq.~{10.2} of Ref.~\cite{privman:91}.  It can be noted that in any
$S = 1/2$ Ising system the infinite-temperature (i.e., independent
spins) limit for the Binder cumulant is $g(0,N) \equiv 1/N$, where N
is the number of spins; as $N = L^{D}$ for a hypercubic lattice, at
infinite temperature $L^{D}g(\tau,L) \equiv 1$. Thus this normalized
Binder cumulant also obeys the high-temperature limit rule for
normalized observables introduced above.

\begin{figure}
  \includegraphics[width=3.4in]{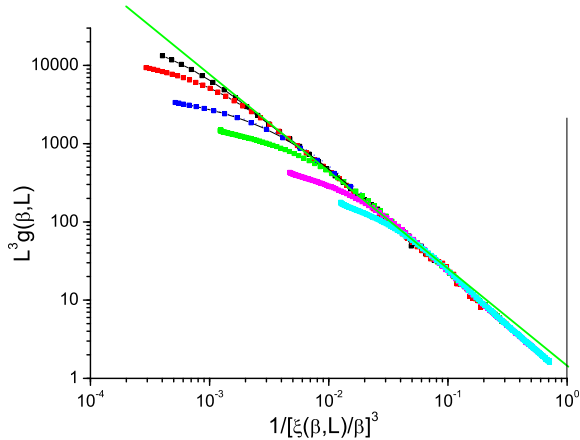}
  \caption{(Color on line) Dimension 3 simple cubic bimodal ISG
    model. Normalized Binder cumulant $L^{3}g(\beta,L)$ against
    reduced correlation length to the power 3,
    $1/(\xi(\beta,L)/\beta)^3$. Sample sizes : $L=32$, $24$, $16$,
    $12$, $8$, $6$ (top to bottom). Green straight line : slope
    $-1.27$.}\protect\label{fig2}
\end{figure}

For Ising models in the thermodynamic limit ThL (bulk or $L \gg
\xi(\tau)$) regime, assuming hyperscaling the critical exponent for
the second field derivative of the susceptibility $\chi_{4}(\beta)$
is~\cite{butera:02}
\begin{equation}
  \gamma_{4}=\gamma +2\Delta = D\nu + 2\gamma
  \label{gam4}
\end{equation}
Thus the bulk $\chi_{4}(\beta)/(2\chi(\beta)^2)$ or the ThL normalized
Binder cumulant $L^D g(\beta,L)$ scales with the critical exponent
$(\nu D + 2\gamma) - 2\gamma = \nu D$, together with appropriate
Wegner correction terms.

\begin{figure}
  \includegraphics[width=3.4in]{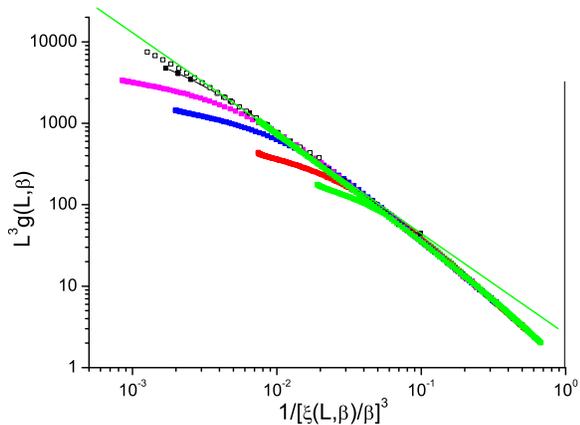}
  \caption{(Color on line) Dimension 3 simple cubic Gaussian ISG
    model. Normalized Binder cumulant $L^{3}g(\beta,L)$ against
    reduced correlation length to the power 3,
    $1/(\xi(\beta,L)/\beta)^3$. Sample sizes : $L=32$, $24$, $16$,
    $12$, $8$, $6$ (top to bottom). Green straight line : slope
    $-1.23$.}\protect\label{fig3}
\end{figure}

The standard ``dimensionless renormalized coupling constant'' can be
defined as
\begin{equation}
  G_{4}(\beta) = \frac{\chi_{4}(\beta)}{\xi(\beta)^{D}\chi(\beta)^{2}} =
  \frac{L^{D}g(\beta,L)}{2 \xi(\beta,L)^{D}}
  \label{G4}
\end{equation}
(other normalizations are also used \cite{butera:02}).  It should be
noted that even in the case of the canonical 2D Ising model
\cite{salas:00} the infinite-$L$ value of $G_{4}(\beta_{c})$ at
criticality depends strongly on the order in which the limits are
taken : the ThL limit $[L \to \infty, \beta \to \beta_{c}]$ or the FSS
limit $[\beta \to \beta_{c}, L \to \infty]$. Also, in order for the
infinite-$L$ Ising $G_{4}$ to become regular analytic to $\beta=0$ the
normalized Ising form $(\beta_{c}/\beta)^{D/2} G_{4}(\beta)$ should be
used, see Ref.~\cite{butera:02} Eq.~(42). This modification is
strictly equivalent to replacing in Eq.~\eqref{G4} $\xi(\beta,L)$ by
$\xi(\beta,L)/\beta^{1/2}$ which is the reduced correlation length
introduced above.

\begin{figure}
  \includegraphics[width=3.4in]{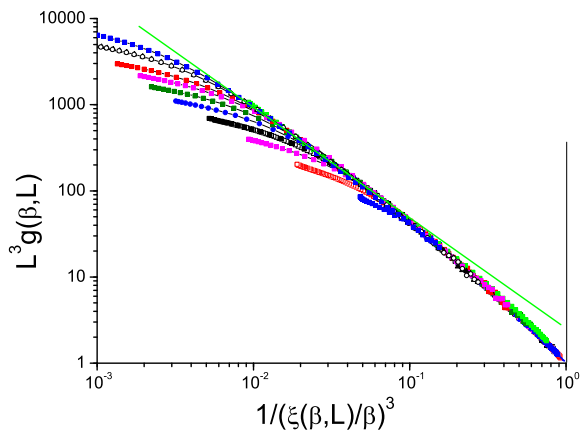}
  \caption{(Color on line) Dimension 3 face-centered cubic Laplacian
    ISG model. Normalized Binder cumulant $L^{3}g(\beta,L)$ against
    reduced correlation length to the power 3,
    $1/(\xi(\beta,L)/\beta)^3$. Sample sizes : $L=28$, $24$, $20$,
    $18$, $16$, $14$, $12$, $10$, $8$, $6$ (top to bottom). Green
    straight line : slope $-1.27$.}\protect\label{fig4}
\end{figure}

From the hyperscaling rule, the critical exponents for
$L^{D}g(\tau,L)$ and $(\xi(\tau,L)/\beta^{1/2})^{D}$ are both
$D\nu$. Fig.~\ref{fig1} shows a plot of $L^{3}g(\tau,L)$ against
$(\xi(\tau,L)/\beta^{1/2})^{-3}$ for the 3D simple cubic $S=1/2$ Ising
model (see Ref.~\cite{campbell:11} for details of the simlations). It
can be seen that within the statistics the ThL data are consistent
with a limiting critical slope $\equiv -1$ and an intercept
$G_{4}\beta_{c}^{3/2}/2 \sim 1.23$, with corrections coming into play
at high temperatures, in full agreement with the hyperscaling rule. It
should be noted that in this form of plot neither the critical inverse
temperature value $\beta_{c}$ nor the critical exponent value $\nu$
need to be introduced.

\begin{figure}
  \includegraphics[width=3.4in]{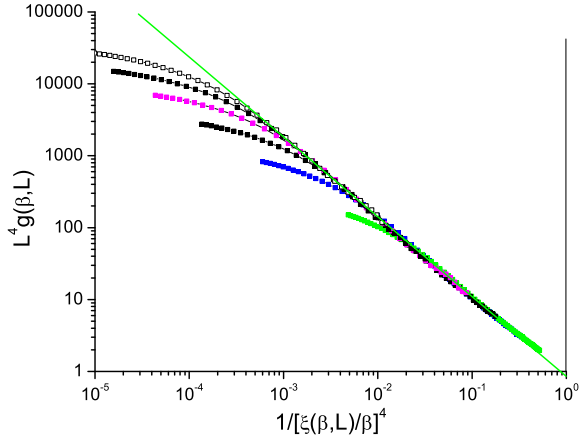}
  \caption{(Color on line) Dimension 4 hypercubic bimodal ISG
    model. Normalized Binder cumulant $L^{4}g(\beta,L)$ against
    reduced correlation length to the power 4,
    $1/(\xi(\beta,L)/\beta)^4$. Sample sizes : $L=14$, $12$, $10$,
    $8$, $6$, $4$ (top to bottom). Green straight line : slope
    $-1.12$.}\protect\label{fig5}
\end{figure}

For the 5D Ising model, in an equivalent ThL plot covering the entire
paramagnetic temperature regime the breakdown of standard hyperscaling
leads to a critical exponent for the normalized Binder cumulant which
is not $D\nu = 5/2$ but $2$ \cite{lundow:18}, i.e., $(D-\theta)\nu$
with a hyperscaling violation exponent $\theta=1$ as discussed above.

\section{Ising spin glasses}\label{sec:IV}
The normalized Binder cumulant against reduced correlation length to
the power $D$ for ISGs can be displayed in just the same way as for
the Ising model data in Fig.~\ref{fig1}. We show in Figs.~\ref{fig2}
and \ref{fig3} data for the standard bimodal and Gaussian interaction
ISGs on the 3D simple cubic lattice, and in Fig.~\ref{fig4} for a more
exotic model, the Laplacian interaction ISG on a face-centered cubic
lattice.  

\begin{figure}
  \includegraphics[width=3.4in]{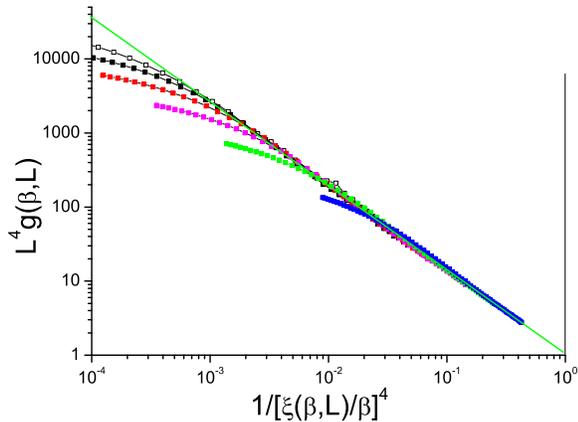}
  \caption{(Color on line) Dimension 4 hypercubic Gaussian ISG
    model. Normalized Binder cumulant $L^{4}g(\beta,L)$ against
    reduced correlation length to the power 4,
    $1/(\xi(\beta,L)/\beta)^4$. Sample sizes : $L=14$, $12$, $10$,
    $8$, $6$, $4$ (top to bottom). Green straight line : slope
    $-1.13$.}\protect\label{fig6}
\end{figure}

In each case the ThL data can be seen to have a critical limit
constant slope with corrections coming into play at high
temperatures. The value of the limit slope is in each case distinctly
stronger than the standard hyperscaling value $-1$, demonstrating that
$(\xi(\beta,L)/\beta)^3)$ and $L^3g(\beta,L)$ have different critical
exponents; there is a violation of hyperscaling.  With the same
formalism as above, the value of the slope in each Figure can be taken
to be equal to $-(D-\theta)/D = -(1-\theta/D)$ with $\theta \sim
-0.80$, $-0.70$ and $-0.80$ respectively for the three 3D ISG models.

\begin{figure}
  \includegraphics[width=3.4in]{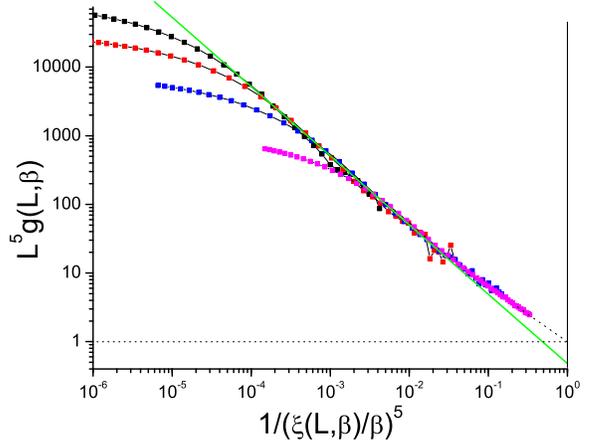}
  \caption{(Color on line) Dimension 5 hypercubic bimodal ISG
    model. Normalized Binder cumulant $L^{5}g(\beta,L)$ against
    reduced correlation length to the power 5,
    $1/(\xi(\beta,L)/\beta)^5$. Sample sizes : $L= 10$, $8$, $6$, $4$
    (top to bottom). Green straight line : slope $-1.00$.}
  \protect\label{fig7}
\end{figure}

\begin{figure}
  \includegraphics[width=3.4in]{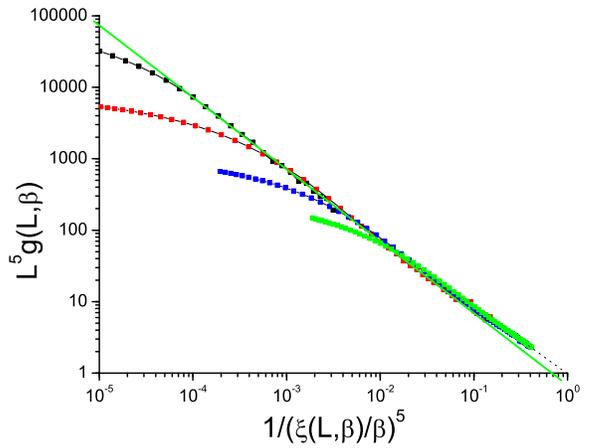}
  \caption{(Color on line) Dimension 5 hypercubic Gaussian ISG
    model. Normalized Binder cumulant $L^{5}g(\beta,L)$ against
    reduced correlation length to the power 5,
    $1/(\xi(\beta,L)/\beta)^5$. Sample sizes : $L= 10$, $8$, $6$, $4$
    (top to bottom). Green straight line : slope $-1.00$.}
  \protect\label{fig8}
\end{figure}

Data for the bimodal and Gaussian 4D models, Figs.~\ref{fig5} and
\ref{fig6}, also show hyperscaling violations with violation exponents
$\theta \sim -0.48$ and $-0.52$ respectively, rather weaker than in
3D.  For the bimodal and Gaussian 5D models, Figs.~\ref{fig7} and
\ref{fig8}, the limiting slopes are close to $-1$ and any violation of
hyperscaling is too weak to be observed.

\begin{figure}
  \includegraphics[width=3.4in]{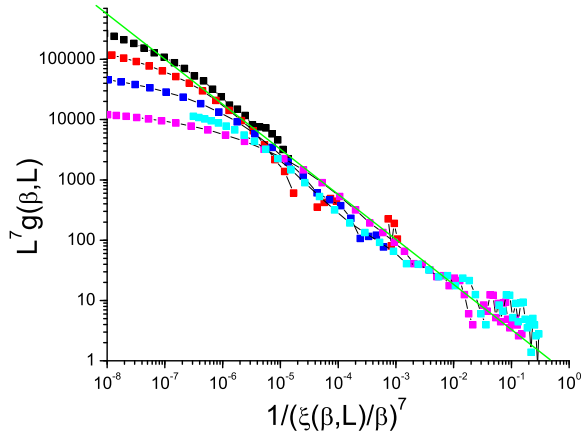}
  \caption{(Color on line) Dimension 7 hypercubic bimodal ISG
    model. Normalized Binder cumulant $L^{7}g(\beta,L)$ against
    reduced correlation length to the power 7,
    $1/(\xi(\beta,L)/\beta)^7$. Sample sizes : $L= 7$, $6$, $5$, $4$,
    $3$ (top to bottom). Green straight line : slope $-0.75$.}
  \protect\label{fig9}
\end{figure}

Finally in the bimodal 7D ISG (above the ISG critical dimension $D=6$)
the equivalent plot, Fig.~\ref{fig9}, is much more noisy than for the
lower dimensions simply because by this dimension the number of spins
in each individual sample becomes very large leading to practical
limitations, particularly for the Binder cumulant. Nevertheless the
slope of the plot can be seen to be lower than $-1$ so as for the 5D
ISG model the violation exponent $\theta \sim 1.75$ is positive.

\section{Conclusion}\label{sec:V}
For the canonical simple cubic Ising model, from data presented in the
form of the normalized Binder cumulant $L^{3}g(\tau,L)$ against the
reduced correlation length to power $D$, $[\xi/\beta^{1/2}]^{D}$ , the
observed critical limit scaling is fully consistent with the expected
standard hyperscaling log-log slope of $-1$ (plus mild corrections at
high temperatures). The equivalent data plots for ISG models in
various dimensions show violations of hyperscaling with violation
exponents which evolve regularly with dimension from strongly negative
for $D=3$ to strongly positive for $D=7$, passing though zero near the upper
critical dimension.

\begin{acknowledgments}
  We would like to thank A.~Aharony, P.~Butera and R. Kenna for
  helpful comments, and H. Katzgraber and K. Hukushima for access to
  their raw numerical data.  The computations were performed on
  resources provided by the Swedish National Infrastructure for
  Computing (SNIC) at the Chalmers Centre for Computational Science
  and Engineering (C3SE).
\end{acknowledgments}


\end{document}